\documentclass[prl,twocolumn,amssymb]{revtex4}
\newcommand{\be}{\begin{equation}}
\newcommand{\ee}{\end{equation}}
\newcommand{\bea}{\begin{eqnarray}}
\newcommand{\eea}{\end{eqnarray}}
\usepackage{epsfig}
\usepackage{graphicx}

\begin{document}

 \title{Trapped Fermi Gases in Rotating Optical Lattices: Realization and Detection of the Topological Hofstadter Insulator}

\author{R. O. Umucal\i lar$^{1}$}
\author{Hui Zhai$^{2,3}$}
\author{M. \"O. Oktel$^{1}$}
\email{oktel@fen.bilkent.edu.tr}

\affiliation{$1$ Department of Physics, Bilkent University, 06800
Ankara, Turkey}
\affiliation{$^2$ Department of Physics,
University of California, Berkeley, California, 94720, USA}
\affiliation{$^3$ Materials Sciences Division, Lawrence Berkeley
National Laboratory, Berkeley, California, 94720, USA}

\date{\today}

\begin{abstract}
We consider a gas of non-interacting spinless fermions in a
rotating optical lattice and calculate the density profile of the
gas in an external confinement potential. The density profile
exhibits distinct plateaus, which correspond to gaps in the single
particle spectrum known as the Hofstadter butterfly. The plateaus
result from insulating behavior whenever the Fermi energy lies
within a gap. We discuss the necessary conditions to realize the
Hofstadter insulator in a cold atom setup and show how the
quantized Hall conductance can be measured from density profiles
using the St\v{r}eda formula.
\end{abstract}
 \maketitle

 Recently, some
fundamental models of many-particle quantum systems have been
experimentally realized using trapped ultra-cold fermions. Some of
these experiments, such as transport in optical lattices
\cite{Inguscio}, has demonstrated well known effects with improved
precision, while others, such as spin imbalanced superfluidity
\cite{Ketterle,Hulet}, have provided access to previously
unexplored regimes. With constantly improving experimental control
over ultra-cold systems, it is expected that many other
fundamental ideas can be tested in laboratory for the first time.

A basic problem in quantum mechanics is the dynamics of a charged
particle moving in a periodic potential under a magnetic field.
The single particle spectrum depends sensitively on the ratio of
the flux through a unit cell of the lattice to flux quantum.  For
a tight binding lattice, a single band splits into narrow magnetic
bands, forming a self-similar energy spectrum known as the
Hofstadter butterfly \cite{Hofstadter}. The gaps in the Hofstadter
spectrum form continuous regions for a finite range of flux. For a
system of non-interacting fermions, it was shown by Thouless
\textit{et. al.} that whenever the Fermi energy lies in one of
these gaps, the Hall conductance of the system is quantized
\cite{Thouless}. This quantization is topological in nature, and
the quantized Hall conductance is determined uniquely by the
magnetic translation symmetry \cite{Dana}. This Hofstadter
insulating phase is a topological insulator that can be
characterized by the first Chern number.

Despite its mathematical elegance, the Hofstadter insulator can
hardly be achieved in solid state systems because the magnetic
field needs to be thousands of Tesla in order to create a magnetic
flux which can be comparable to one flux quantum per unit cell
\cite{magnetic-field}. While in some experiments super-lattice
structures have been used to study the splitting of Landau levels
under a periodic potential, the tight-binding regime (strong
magnetic field limit) has never been experimentally realized
\cite{Klitzing}. The main purpose of this Letter is to propose an
alternative way to achieve and to experimentally study this
topological insulator by using ultra-cold Fermi gases in a
rotating optical lattice. We discuss (i) the conditions to realize
the Hofstadter insulator in rotating optical lattices, and (ii)
the manifestation of this insulator in real space density profile
and the method to detect the Hall conductance in a cold atom
setup.

The Hamiltonian for a particle in a rotating lattice is
\begin{eqnarray}\label{Hamiltonian}
    H &=& \frac{1}{2m}\mathbf{p}_{\bot}^2 +
    \frac{1}{2}m\omega_{\bot}^2r^2 - \Omega \mathbf{\hat{z}}\cdot
    \mathbf{r}\times \nonumber
    \mathbf{p}_{\bot} \\ &+& V_0\big[\sin^2(k x)+\sin^2(k y)\big],
\end{eqnarray}
where $m$ is the mass of the particle, $\Omega$ is the rotation
frequency, and $\omega_{\bot}$ is the transverse trapping
frequency. $\mathbf{p}_{\bot} = (p_x,p_y)$, $\mathbf{r} = (x,y)$,
and $V_0$ is the maximum depth of the optical potential created by
a laser light with wave number  $k = 2\pi/\lambda$ (for
counter-propagating laser beams lattice constant $a$ is equal to
$\lambda/2$). Photon recoil energy $E_{\text{R}}$, defined as $\hbar^2
k^2/(2m)$ is introduced as the energy unit in the following
discussion. This Hamiltonian can be rearranged as
\begin{eqnarray}\label{Hamiltonian2}
    H &=& \frac{(\mathbf{p}_{\bot}-m\Omega \mathbf{\hat{z}}\times \mathbf{r})^2}{2m}
    +\nonumber
    V_0\big[\sin^2(k x)+\sin^2(k y)\big] \\ &+&
    \frac{1}{2}m(\omega_{\bot}^2-\Omega^2)r^2.
\end{eqnarray}
When $\Omega$ is close to $\omega_\bot$, the third term represents
a smooth potential in space. The first term describes the motion
of a particle under a perpendicular magnetic field with strength
$B = 2mc\Omega/e$. This effective magnetic field description has
led a number of authors to study the quantum-Hall type of physics
in rotating quantum gases \cite{QH}. Here, we consider the
presence of both rotation and a lattice potential
\cite{latticeQH}, employing the Hofstadter model, which is
appropriate for the description of a non-interacting gas of
fermions in a lattice.

We now discuss the conditions for simulating the Hofstadter model
in an ultra-cold optical lattice system:

(1) Hofstadter model is expressly single band; the motion of
fermions in the periodic potential has to be well described by a
single-band tight-binding model with nearest neighbor hopping.
This requirement means that there has to be a finite gap between
$s$- and $p$-bands, and the dispersion of the $s$-band can be well
approximated by a cosine function. From the band structure
calculations for optical lattice potentials, one can easily show
that this requirement is fulfilled when $V_0>3E_{\text{R}}$
\cite{Ref}. A rotating lattice experiment has recently been
carried out \cite{Cornell}. Although in this particular experiment
$V_0<1E_{\text{R}}$, which is not deep enough to reach the tight
binding regime, there is no fundamental reason against increasing
$V_0$ a few more $E_{\text{R}}$, as it has been routinely done in
static lattice experiments \cite{Bloch-Nature}.

(2) The ``magnetic field" has to be strong enough. The
dimensionless parameter $\phi = a^2 B /(hc/e)$ is the magnetic
flux quantum per plaquette and is connected to the rotation
frequency $\Omega$ as $\phi = 2ma^2\Omega/h$.  For rotation to
create an effective magnetic flux close to one flux quantum, the
rotation frequency $\Omega$ must be close to $E_{\text{R}}/\hbar$ of the
lattice. In typical optical lattice experiments the recoil
frequency is a few kHz, thus, rotation of the lattice at hundreds
to thousands of Hz would be enough to reach the high magnetic
field limit.

(3) To observe the insulating behavior, the temperature has to be
lower than the gap of the insulator. The gap for the Hofstadter
insulator is comparable to the hopping amplitude $t$. In a
moderately deep lattice, for which $V_0 \sim 3-7E_{R}$, for
instance, $t$ is of the order of $1-10 nK$. This is below
currently attainable temperatures. However, cooling fermions to
this regime is not more difficult than achieving degenerate Fermi
gases in a lattice, and the latter is now the major goal pursued
in many key laboratories in this field.

(4) The Hofstadter model is a non-interacting one. Strong
repulsive or attractive interactions can lead to either exciton
instability or BCS instability of the insulating phase (which will
be discussed elsewhere), and therefore diminish its topological
behavior. In ultra-cold atom experiments a single species of
fermions is naturally non-interacting due to the Pauli exclusion
principle.

Even when the above conditions are satisfied, there are two
factors which complicate the correspondence between the results of
ultra-cold atom experiments and theories developed for `bulk'
systems.  The first is the presence of an external confining
potential in all ultra-cold atom experiments, i.e. the third term
in Eq.(\ref{Hamiltonian2}). The second factor is that transport
measurements are usually very hard for cold atom systems.
Hereafter, we first calculate the density profile of a
non-interacting Fermi gas in a rotating optical lattice, in the
presence of a smooth external potential \cite{Ho}. We find that
the presence of the residual trapping potential does not preclude
the observation of the effects of single particle spectrum, as
long as it is varying smooth enough on the lattice length scale.
Secondly, we show that one of the most important transport
properties, namely the Hall conductance which reflects the
topological nature, can be inferred from the measurement of the
density profile due to the well known St\v{r}eda formula
\cite{Streda}.

When the residual trapping potential is slowly varying, we can
utilize the local-density approximation (LDA) in which we define a
local chemical potential $\mu_l(r)$ (or Fermi energy) as
\begin{eqnarray}\label{chemical}
\mu_l({\bf r}) = \mu-V({\bf r}),
\end{eqnarray}
where in our case $V(r) = m(\omega_{\bot}^2-\Omega^2)r^2/2$. In
what follows, we shall denote $(\omega_{\bot}^2-\Omega^2)$ by
$\omega^2$.

\begin{figure}
\includegraphics[scale=0.22]{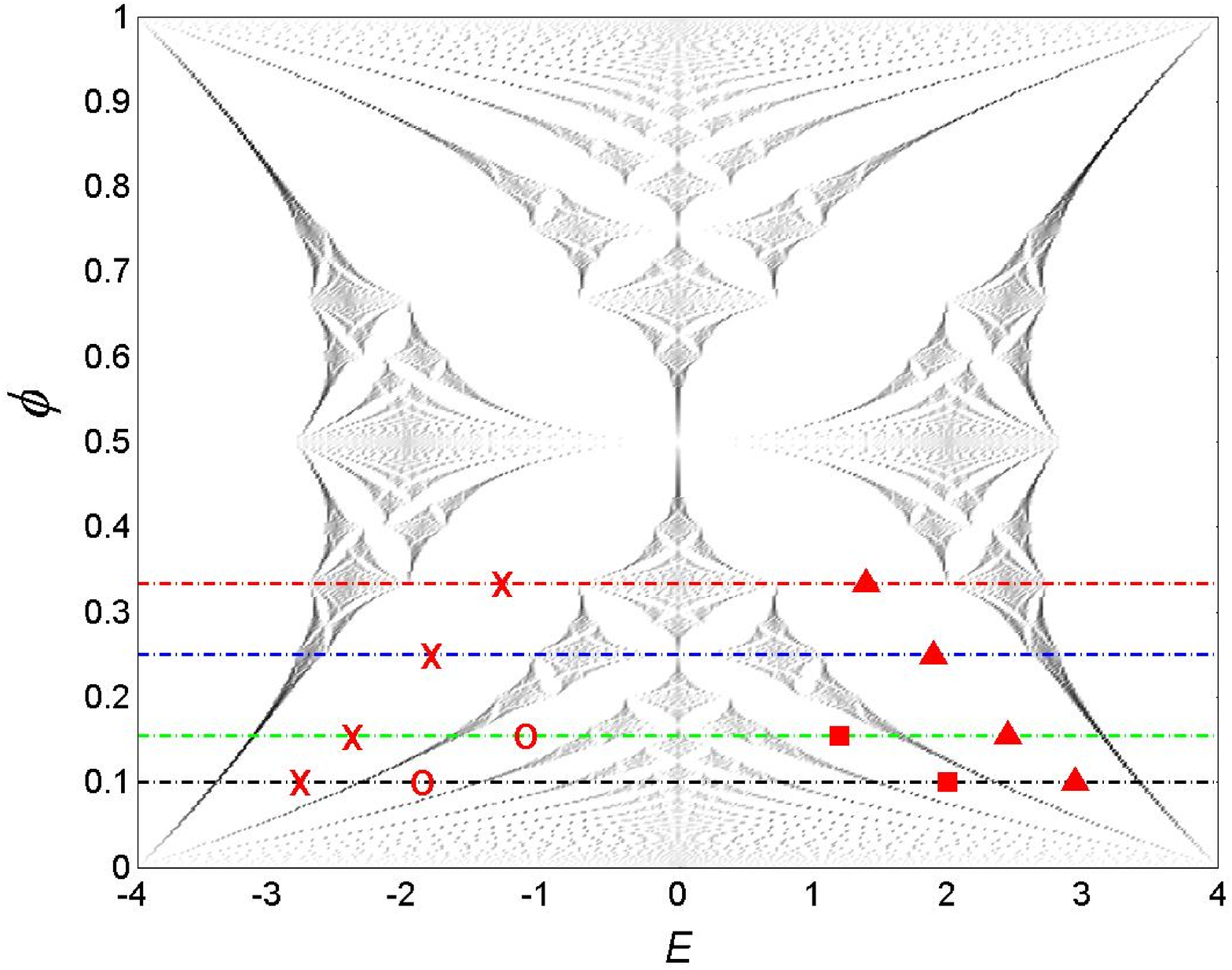}
\caption{Density of states for the Hofstadter butterfly. Darker
regions imply greater density. Dashed lines represent the
trajectory of local Fermi energy from the center to the edge of
the cloud, for different values of $\phi$ corresponding to those
used in Fig. \ref{fig:phi100_134over401}, namely $\phi = 1/3, 1/4,
1/7$, and $1/10$. Regions marked by $\times$ and $\blacktriangle$
have Hall conductance $\sigma_{xy}=\pm 1$; and marked by $\circ$
and {\tiny $\blacksquare$} have $\sigma_{xy}=\pm 2$. }
\label{fig:Hofsdens}
\end{figure}

To find the density profile $n(r)$, we simply count the number of
states below $\mu_l(r)$ for the corresponding uniform system as
\begin{eqnarray}\label{LDA}
n(r)=\int d\epsilon D(\epsilon)\Theta(\mu_l(r)-\epsilon).
\end{eqnarray}
So, we shall first find the density of states $D(\epsilon)$ for
non-interacting fermions in a periodic lattice under a magnetic
field. By using the Landau gauge $\overrightarrow{A} = Bx
\hat{y}$, one can construct the Hamiltonian for this system as
follows
\begin{eqnarray}\label{Hamiltoniantight}
H = -t\sum_{\langle i,j\rangle}a_i^\dag a_j e^{i A_{ij}},
\end{eqnarray}
where $a_i$ ($a_i^\dag$) is the fermionic annihilation (creation)
operator at site $i$ and the sum is over nearest neighbor sites.
Magnetic field affects the Hamiltonian through $A_{ij}$, which is
equal to $\pm 2\pi n \phi$ , if $i$ and $j$ have the same $x$
coordinate $na$, and is $0$ otherwise, with the sign being
determined by the hopping direction. The hopping strength $t$ can
be obtained as a function of $V_0/E_R$ from a simple
band-structure calculation. The density of states is displayed in
Fig. \ref{fig:Hofsdens}. When $\phi$ is a rational number $p/q$,
with $p$ and $q$ being relatively prime integers, the energy band
divides into $q$ bands \cite{Hofstadter}. In this spectrum the
gaps form continuous regions in $\phi-E$ plane, although the band
edges are fractal. When $\mu$ lies in a gap, the system is an
insulator, and as one changes $\phi$ and $\mu$, the topological
nature and the Hall conductance of the insulator do not change as
long as one remains within the same gap \cite{Thouless}. The
largest two gaps correspond to insulators with Hall conductance
$\sigma_{xy}=\pm1$, and the second largest ones have Hall
conductance $\sigma_{xy}=\pm 2$ and so on, as marked in Fig.
\ref{fig:Hofsdens}.

To calculate the integral in Eq. (\ref{LDA}) efficiently, we note
that if we take $q$ to be sufficiently large, the band-widths of
each sub-band become negligibly small. Counting the number of
states then reduces to counting these bands in certain intervals.
In our numerical procedure, we determine the sub-bands with their
edge values $\varepsilon_{edge}$. So the number of states per
plaquette in the two dimensional case can conveniently be
expressed as
\begin{eqnarray}\label{2dnumberofstates}
n(\mu) =
\frac{1}{2q}\sum_{\varepsilon_{edge}}\Theta(\mu-\varepsilon_{edge}).
\end{eqnarray}
In all of our calculations we took $q = 401$, which is a prime
number allowing $p$ to be successive integers. $\phi$ values for
other small denominators of $q$ are approximated by properly
choosing $p$. For instance, $\phi=1/10$ is approximated by
$40/401$, $1/4$ by $100/401$ and $1/3$ by $134/401$.

\begin{figure}
\includegraphics[scale=0.4]{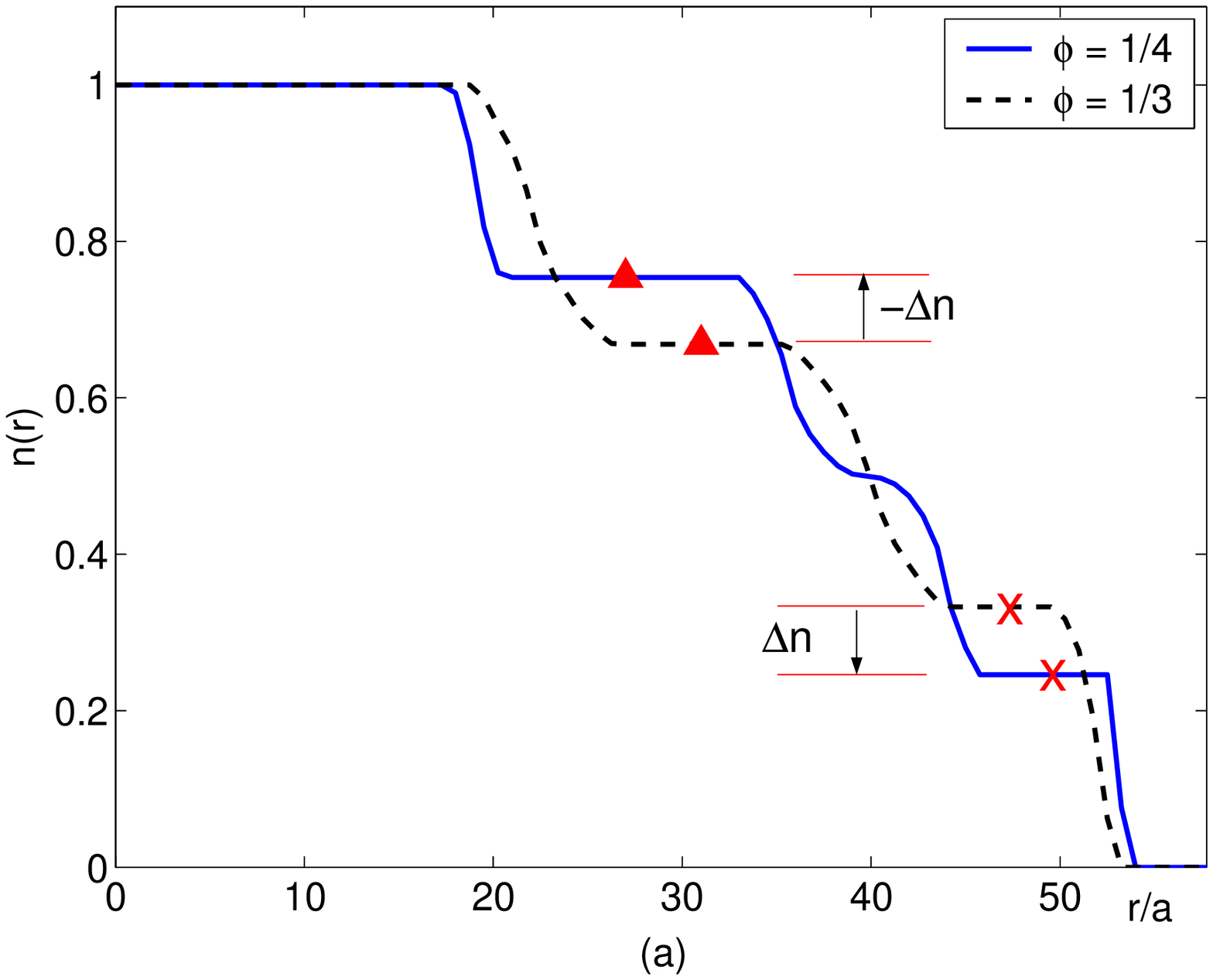}
\includegraphics[scale=0.4]{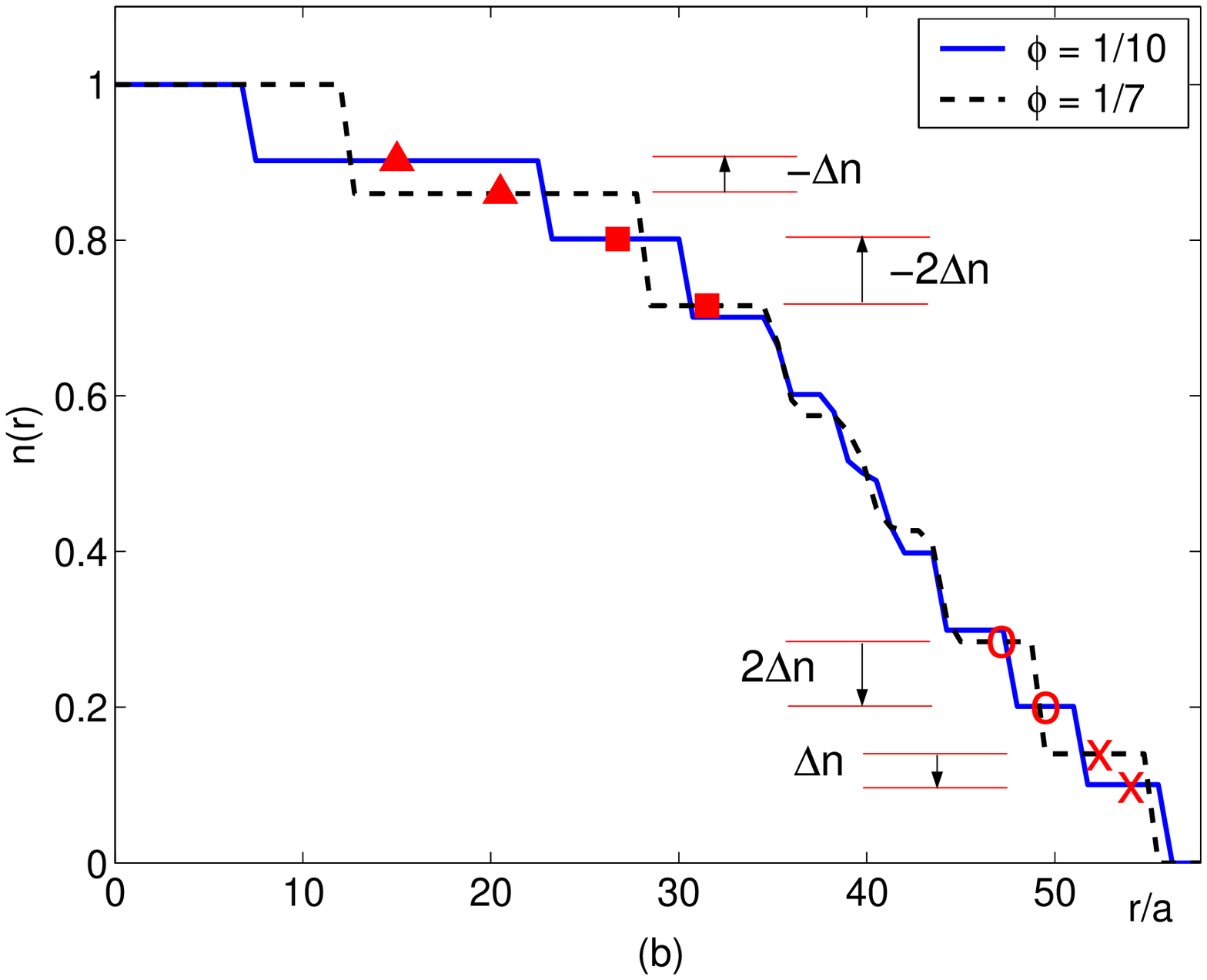}
\caption{(a) Density profiles for 5000 fermions with $\phi = 1/4$,
$\Omega = 7.2992$ kHz, $\omega_{\bot} = 7.3078$ kHz (solid line)
and $\phi = 1/3$, $\Omega = 9.7809$ kHz, $\omega_{\bot} = 9.7873$
kHz (dashed line). (b) Density profiles for 5000 fermions with
$\phi=1/10$, $\Omega = 2.9197$ kHz, $\omega_{\bot} = 2.9412$ kHz
(solid line) and $\phi=1/7$, $\Omega = 4.1605$ kHz, $\omega_{\bot}
= 4.1756$ kHz (dashed line). Length is measured in units of
lattice constant $a$.  Density is given in units of particles per
lattice site.}\label{fig:phi100_134over401}
\end{figure}

\begin{figure}
\includegraphics[scale=0.4]{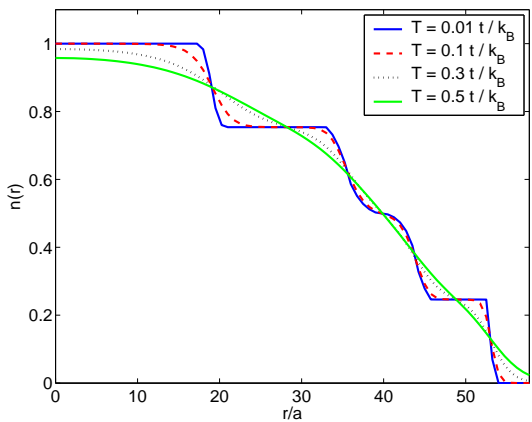}
\caption{(Color online) Density profile for 5000 fermions at
several temperatures when $\phi = 1/4$.
Plateaus become indiscernible when $k_B T \sim 0.5 t$.}
\label{fig:phi100over401T}
\end{figure}

We now present the density profiles for several $\phi$ values. To
make a connection with experiments, we refer to the work in Ref.
\cite{Esslinger} in which $^{40}\textrm{K}$ atoms are stored in an
optical lattice with lattice constant $a = 413$ nm. We take $V_0 =
5 E_{\text{R}}$, which gives $t = 0.066 E_{\text{R}}$. The
parameters at hand yield $E_{\text{R}}/\hbar = 45.98$ kHz and
$t/\hbar = 3.035$ kHz. With the choice $\omega \sim 355$ Hz, the
gas extends over approximately $60$ lattice sites in the radial
direction, so that the assumption of LDA is satisfied. In Figs.
\ref{fig:phi100_134over401} and \ref{fig:phi100over401T} we fixed
the number of fermions at $5000$.

When the local chemical potential $\mu_{l}(r)$ lies in one of the
gaps, we have $\partial n(r)/\partial \mu(r)=0$ because of
vanishing compressibility. Hence, as one can see by comparing Fig.
\ref{fig:Hofsdens} and Fig. \ref{fig:phi100_134over401},
corresponding to the energy gaps in the single particle spectrum,
there appear plateaus in the density profile. The discernible
number of plateaus is related to the size of the energy gaps. For
instance, in Fig. \ref{fig:phi100_134over401}(a), the plateau with
$n=1$ is the band insulator with completely filled band, which is
topologically trivial and has vanishing Hall conductance. Apart
from that, for $\phi=1/3$, the chemical potential trajectory
passes through two gap regions which gives two plateaus with $n=
0.333$ and $n=0.667$ respectively. While for $\phi=1/4$, there are
totally four sub-bands, but two of them touch at $\mu=0$, so there
are also two gap regions corresponding to two plateaus with $n=
0.25$ and $n=0.75$. In Fig. \ref{fig:phi100_134over401}(b) we
choose two $\phi$'s with larger $q$, where there are more gaps in
the spectrum and therefore more density plateaus. Experimentally,
the smaller gap one wants to find, the more difficult it is,
because it requires larger system size and lower temperature.

In Fig. \ref{fig:phi100over401T} we show the temperature effect on the
visibility of plateaus. We
implement the effect of finite temperature by incorporating the
Fermi-Dirac distribution into our calculations as
\begin{eqnarray}\label{2dnumberofstates_finiteT}
n_{2D}(\mu_l(r), T) = \frac{1}{2q}\sum_{\varepsilon_{edge}}
\frac{1}{\exp[(\varepsilon_{edge}-\mu_l(r))/k_B T]+1}.
\end{eqnarray}
We observe from Fig. \ref{fig:phi100over401T} that plateaus will
be smeared out when $k_B T > 0.5 t$.

As shown by Thouless \textit{et. al.}, the topological distinction
of those insulators manifests itself in the Hall conductance,
which should be quantized in units of $e^2/h$ \cite{Thouless}.
Here, we propose a method to read out the information about the
Hall conductance quantization from the obtained density profiles.
One can observe how the particle density within a particular
plateau (labelled by the corresponding gap) changes with the
rotation frequency, and the change of plateau density with respect
to rotation frequency is directly related to the Hall conductance
of that plateau as follows.

St\v{r}eda obtained a formula for the Hall conductance of a two
dimensional charged system as \be \sigma_{xy}= e c \frac{\partial
N}{\partial B}, \ee which is valid when the Fermi energy lies in a
gap \cite{Streda}. Here $N$ is the number of levels below the
Fermi energy.  For a neutral gas we can define a similar response
function $ \gamma_{xy}  = J_x / F_y $,  where $J_x$ is the mass
current in $x$ direction induced by a force $F_y$ in the $y$
direction. Then the St\v{r}eda formula for the rotating system can
be written as \be \gamma_{xy}= \frac{1}{2} \frac{\partial
N}{\partial \Omega}. \ee Expressing the rotation frequency in
terms of $\phi$ and $N$ in terms of the density per plaquette
$n(r)$ (as in our density plots), one obtains \be \gamma_{xy}=
\frac{m}{h} \frac{\partial n}{\partial \phi}. \ee

To measure Hall conductance, we first choose two $\phi$ values,
and identify the plateaus in both density profiles that correspond
to the same gap. The density difference of those two plateaus
divided by the difference between $\phi$ values gives the Hall
conductance. We use Fig. \ref{fig:phi100_134over401} as an example
to show how this procedure works. The plateaus are marked by the
same symbol as their corresponding gaps in Fig.
\ref{fig:Hofsdens}. Two plateaus marked by the same symbol are the
same insulating phase. In Fig. \ref{fig:phi100_134over401}(a), for
that marked by $\times$, $\Delta n=0.333-0.25$, and $\Delta
n/\Delta\phi=1$, for that marked by $\blacktriangle$, $\Delta
n=0.667-0.75$, and $\Delta n/\Delta\phi=-1$. In Fig.
\ref{fig:phi100_134over401}(b), one can get the same quantization
number for the plateaus marked by $\times$ and $\blacktriangle$.
In addition, there are more plateaus, such as those marked by
$\circ$ and {\tiny $\blacksquare$} corresponding to the second
largest gaps. For that marked by $\circ$, $\Delta n=0.284-0.201$
and $\Delta n/\Delta\phi=2$, and for that marked by {\tiny
$\blacksquare$}, $\Delta n=0.716-0.802$ and $\Delta
n/\Delta\phi=-2$.

In summary, we discussed the experimental conditions for
simulating Hofstadter model in rotating optical lattices, such as
lattice depth, rotational frequency and temperature. We calculated
the density profile in the presence of a smooth residual trapping
potential, and showed how the density plateaus reflect the
insulating behavior in a `` magnetic field " with incommensurate
filling number. We also propose a method to measure the Hall
conductance from real space density profiles, without doing
transport experiments.

R.O.U. is supported by T\"{U}B\.{I}TAK. H.Z. and M.\"{O}.O. would
like to thank the Institut Henri Poincar\'{e}'s program on
``Quantum Gases" for hospitality. M.\"{O}.O. is supported by
T\"{U}B\.{I}TAK-KAR\.{I}YER Grant No. 104T165 and a
T{\"U}BA-GEB\.{I}P grant.

\end{document}